\documentclass[reprint,secnumarabic,amssymb,amsmath,aip,jcp,groupedaddress,frontmatterverbose]{revtex4-1}

\usepackage{graphicx}
\usepackage{dcolumn}
\usepackage{bm}
\usepackage{epstopdf}
\usepackage{amsmath,amsbsy,amsfonts,amssymb}
\usepackage{epsfig}
\usepackage{upgreek}

\draft

\begin{document}


\title{The twofold diabatization of the KRb $(1\sim 2)^1\Pi$ complex in the framework of \emph{ab initio} and deperturbation approaches}

\author{S. V. Kozlov, E. A. Pazyuk, A. V. Stolyarov}
\address{Department of Chemistry, Lomonosov Moscow State University, 119991 Moscow, Leninskie gory 1/3, Russia}

\date{\today}

\begin{abstract}
We performed a diabatization of the mutually perturbed $1^1\Pi$ and $2^1\Pi$ states of KRb based on both electronic structure calculation and direct coupled-channel deperturbation analysis of experimental energies. The potential energy curves (PECs) of the diabatic states and their scalar coupling were constructed from the \textit{ab initio} adiabatic PECs by analytically integrating the radial $\langle \psi_1^{ad}|\partial /\partial R|\psi_2^{ad}\rangle$ matrix element obtained by a finite-difference method. The diabatic potentials and electronic coupling function were refined by the least squares fitting of the rovibronic termvalues of the $1^1\Pi\sim 2^1\Pi$ complex. The empirical PECs combined with the coupling function as well as the diabatized spin-orbit coupling and transition dipole matrix elements are useful for further deperturbation treatment of both singlet and triplet states manifold.
\end{abstract}

\maketitle


\section{Introduction}
The accurate representation of the interacting electronic states plays a key role in understanding the detailed mechanism of the photo and collisionally induced chemical reactions. The singlet-triplet levels of alkali metal dimers serve as an intermediate state in the two step optical transformation of the weakly bound atomic pairs into the absolute ground $X^1\Sigma^+(v=0,J=0)$ molecular state~\cite{Kremsbook, pazyuk2015}. To suppress the undesired spontaneous transitions to the low-lying states a coherent stimulated Raman adiabatic passage~\cite{stirap} (STIRAP) is often used.

\begin{figure}[h!]
\centering
  \includegraphics[scale=0.4]{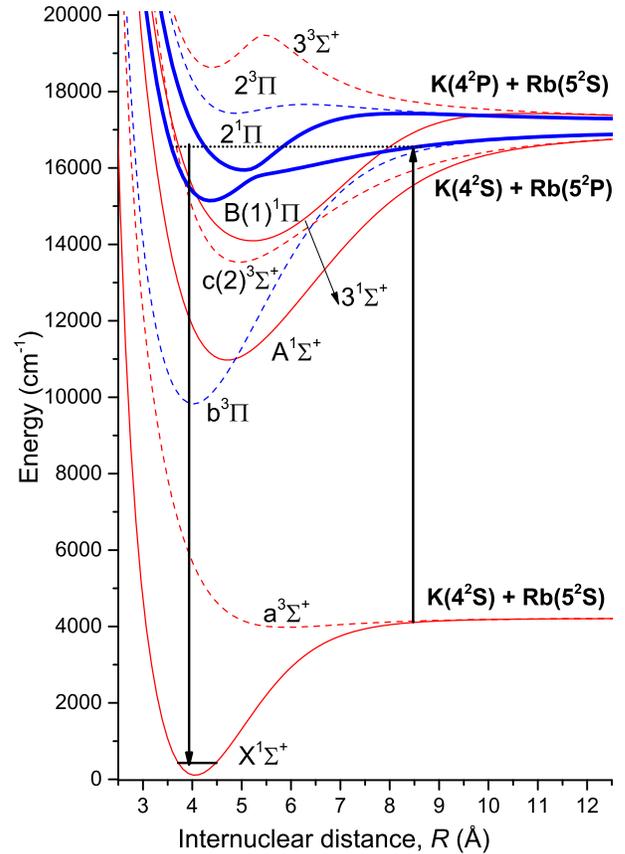}
  \caption{Scheme of the \emph{ab initio} adiabatic potential energy curves~\cite{rousseau2000} of the KRb electronic states correlated to the lowest three dissociation limits. The arrows denote a possible two step stimulated Raman  adiabatic passage~\cite{stirap1}.}
  \label{PECs_KRb}
\end{figure}

The photoassociative production and trapping of ultracold KRb molecules has been performed~\cite{wang2004PRL}. The resonance coupling of the $B(1)^1\Pi$ and $2^1\Pi$ states of KRb (see, Fig.~\ref{PECs_KRb}) is found to be a promising pathway for direct photoassociative formation of the $X(0,0)$ ultracold molecules~\cite{Stwalley2010}. The rigorous multi-channel modeling of the laser formation of vibrationally cold KRb molecules has been accomplished in Ref.~\onlinecite{Kotochigova2009}. The $a^3\Sigma^+\to A^1\Sigma^+\sim b^3\Pi\to X^1\Sigma^+$ and $a^3\Sigma^+\to B^1\Pi \sim c^3\Sigma^+\to X^1\Sigma^+$ optical schemes to create ultracold KRb molecules have been studied~\cite{borsalino2014} by using the \emph{ab initio} potential energy curves (PECs), spin-orbit coupling (SOC) and transition dipole moment (TDM) functions. The combination of a molecular beam (MB) and an ultracold molecule (UM) excitation spectroscopy~\cite{JTKimPhysRev2011} was used to identify the optimal $a^3\Sigma^+(v^{\prime\prime}_a=21)\to B^1\Pi(v_1^{\prime}=8)\to X^1\Sigma^+(v^{\prime\prime}_X=0)$ STIRAP pathway for the $^{39}$K$^{85}$Rb molecule assembling. The magnetoassociated fermion $^{40}$K$^{87}$Rb molecules have been STIRAP transferred~\cite{stirap1, Aikawa2010}  to the lowest $X(0,0)$ level through the $B^1\Pi \sim c^3\Sigma^+$ levels located near the second dissociation threshold.

\begin{figure}[h!]
\centering
\includegraphics[scale=0.4]{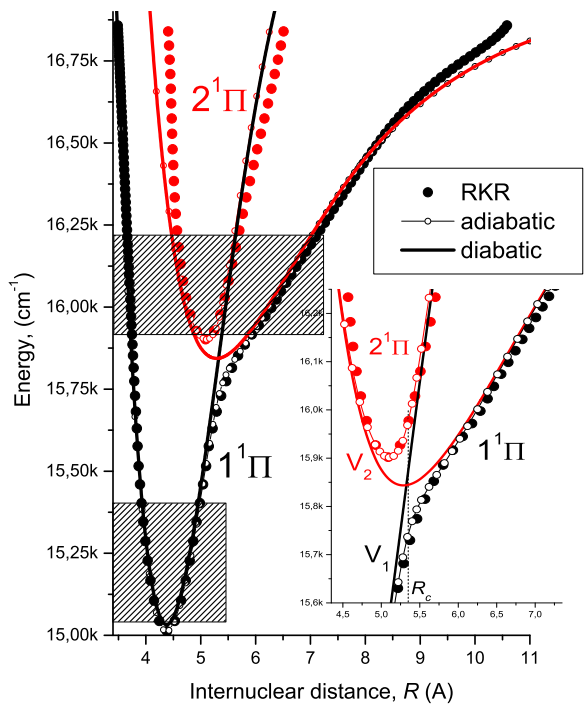}
  \caption{The empirical adiabatic (open symbols - present work, closed symbols - RKR) and diabatic (solid lines) PECs available for the $(1,2)^1\Pi$ twin states of KRb. The RKR potential for the $2^1\Pi$ state was built using the Dunham coefficients~\cite{amiot2000}, while the RKR points of the $B(1)^1\Pi$ state were borrowed from Ref.~\onlinecite{kasahara1999}. The shadowed regions indicate the rovibronic $E^{exp}_{v^{\prime}J^{\prime}}$ termvalues data sets~\cite{amiot2000, amiot2016} included in the present CC deperturbation analysis. The inset enlarges the region in the vicinity of the crossing point of $V_1(R)$ and $V_2(R)$ diabatic PECs.}
  \label{PECs_KRb_emp}
\end{figure}

A comprehensive review of modern spectroscopic studies of the KRb electronic states can be found in the e-book~\cite{Stwalley}. The mutually perturbed $1^1\Pi$ and $2^1\Pi$ states converging to the second K(4$^2$S)+Rb(5$^2$P) and third K(4$^2$P)+Rb(5$^2$S) dissociation thresholds were investigated~\cite{okada1996, kasahara1999} using Doppler-free optical-optical double resonance polarization spectroscopy (OODRPS). In the subsequent laser induced $3^1\Pi\to 2^1\Pi$ fluorescence (LIF) studies~\cite{amiot2000, amiot2016} of both $^{39}$K$^{85}$Rb and $^{39}$K$^{87}$Rb isotopologues by Fourier transform spectroscopy (FTS) the vibrational numbering of the $2^1\Pi(v_2^{\prime})$ state was corrected by 6 vibrational quanta. The experimental rovibronic term values of the $1^1\Pi\sim 2^1\Pi$ complex were reduced to the "effective" band $E_{v^{\prime}}$, $B_{v^{\prime}}$ and conventional Dunham $Y_{ij}$ molecular constants~\cite{kasahara1999, amiot2000}. The Rydberg-Klein-Rees (RKR) potential was constructed for the adiabatic $B(1)^1\Pi$ state (see, Fig.~\ref{PECs_KRb_emp}). The vibrational termvalues of the lowest $B^1\Pi(v_1^{\prime}\in [0,21])$ levels were obtained during the spectroscopic analysis of the MB experiment~\cite{kim2011}. The ground singlet $X^1\Sigma^+$ and triplet $a^3\Sigma^+$ states were comprehensively studied by means of high resolution LIF spectra~\cite{Pashov2007} coming from the spin-orbit coupled $B^1\Pi \sim c^3\Sigma^+$ levels.

The adiabatic potentials, permanent and transition dipole moments for the radially coupled $1^1\Pi$ and $2^1\Pi$ states were first \emph{ab initio} calculated in Refs~\onlinecite{leininger1995, leininger1997}. The comprehensive set of non-relativistic PECs, permanent and transition dipole moments for the ground and excited states of KRb are available~\cite{yiannopoulou1996, park2000, rousseau2000, Beuc2006, Ab_KRb_2016, Minsk2016} as well. The SOC effect has been included into \emph{ab initio} calculations in Refs.~\onlinecite{rousseau2000, Kotochigova2003, Kotochigova2009, Ab_KRb_2016}.

Among other alkali diatomics, the KRb molecule stands out because of the high density of the electronic states belonging to both singlet and triplet manifolds. This is attributed to the accidental close values of the ionization potential, electronic affinity and the polarizability of K and Rb atoms in their ground states as well as the almost degenerate energies of the first excited K(4$^2$P) and Rb(5$^2$P) states~\cite{Radzig, AtomPol, Atom}. The high density of the low-lying covalence and ion-pair states apparently leads to the pronounced radial coupling effect between the states of the same spatial and spin symmetry. This appears (see, for example, Figs.~\ref{PECs_KRb} and \ref{PECs_KRb_emp}) as an avoided crossing of the corresponding adiabatic PECs as well as a sharp dependence of the relevant electronic matrix elements on the internuclear distance $R$. The sharp $R$-dependence of the spin-orbit, angular and radial coupling matrix elements embarrasses a deperturbation analysis while the abrupt $R$-variation of the adiabatic TDM functions prevents to a straightforward simulation of radiative properties.

The electronic coupling matrix element $V_{12}\approx U_1(R_c)-U_2(R_c)$ estimates near the avoided crossing point $R_c$ of the adiabatic PECs warn that a conventional adiabatic approximation is not ideally suitable for representation of the twin $(1,2)^1\Pi$ states of KRb, since the so-called adiabaticity parameter~\cite{field} $\gamma\equiv V_{12}/\sqrt{\omega_1\omega_2}$ is close to $3$. Here, $\omega_i$ are harmonic frequencies of the interacting adiabatic states. To our best knowledge, a global deperturbation analysis of the $1^1\Pi\sim 2^1\Pi$ complex has not been performed yet in the framework of either adiabatic or diabatic approximation.

In the present work, we performed a twofold diabatization of the KRb $1^1\Pi\sim 2^1\Pi$ complex by means of mutually complementary methods, namely: \textit{ab initio} electronic structure calculations and direct coupled-channel (CC) deperturbation treatment of experimental termvalues~\cite{okada1996, kasahara1999, kim2011, Stwalley, amiot2000, amiot2016}.

\section{\emph{Ab initio} diabatization of the $(1\sim 2)^1\Pi$ twin states}

The simplest two-state transformation (diabatization) of the adiabatic electronic wavefunctions $\psi_{1,2}(R)$ to their diabatic counterparts $\varphi_{1,2}(R)$ can be realized by the unitary transformation~\cite{field}
\begin{eqnarray}\label{diabTR}
 \begin{pmatrix} \varphi_1 \\ \varphi_2 \end{pmatrix} =
 \begin{pmatrix} \cos\theta && \sin\theta\\ -\sin\theta && \cos\theta \end{pmatrix}
 \begin{pmatrix} \psi_1 \\ \psi_2 \end{pmatrix}
\end{eqnarray}
where the rotation angle $\theta(R)$ is evaluated as a function of the internuclear distance $R$ by integration of the radial coupling matrix element:
\begin{eqnarray}\label{Rcoup}
  B_{12}\equiv \langle \psi_1 |{\partial }/{\partial R}|\psi_2\rangle = \frac{d \theta}{d R}.
\end{eqnarray}
The integration of the \emph{ab initio} calculated radial coupling matrix element (\ref{Rcoup}) is performed implicitly by means of a smooth interpolation of the original point-wise $B_{12}(R)$ function in the vicinity of a dominant maximum (which is located near the avoided crossing point $R_c$ of the corresponding adiabatic potentials) by the simplest Lorentz form~\cite{field}:
\begin{eqnarray}\label{B12FIT}
B_{12}(R) \approx \frac{w}{4(R-R_c)^2+w^2}
\end{eqnarray}
with the two $R$-independent parameters $R_c$ and $w$. Then, the required rotation angle function
\begin{eqnarray}\label{dangFIT}
 \theta(R)=\frac{1}{2}\arctan \left[\frac{2(R-R_c)}{w} \right] +\frac{\pi}{4}
\end{eqnarray}
is prolonged to the $R\in [0,+\infty)$-range in order to accomplish a diabatization of the corresponding electronic wave functions (\ref{diabTR}).

The diabatic potentials $V_{1,2}(R)$ and electronic coupling matrix element $V_{12}(R)$ are calculated from the adiabatic PECs $U_{1,2}(R)$ via the relations:
\begin{eqnarray}\label{ad_di}
V_1    &=& \cos^2\theta U_1 + \sin^2\theta U_2 \nonumber\\
V_2    &=& \sin^2\theta U_1 + \cos^2\theta U_2 \nonumber\\
V_{12} &=& \sin 2\theta |U_1-U_2|/2
\end{eqnarray}

The adiabatic PECs $U^{ab}_i(R)$ were evaluated for the low-lying excited $(1-3)^{1,3}\Sigma^+$ and $(1,2)^{1,3}\Pi$ states converging to the lowest three dissociation limits (see Fig.~\ref{PECs_KRb}) in the basis of the zeroth-order (spin-averaged) electronic wavefunctions corresponding to pure (\textbf{a}) Hund's coupling case. To diminish the systematic $R$-depended error (first of all, basis set superposition error) the originally calculated adiabatic potentials $U^{ab}_i$ for the excited states were corrected due to the semi-empirical relation~\cite{pazyuk2015}:
\begin{eqnarray}\label{difbased}
U_i(R) = [U^{ab}_i(R)-U_X^{ab}(R)] + U_X^{emp}(R)
\end{eqnarray}
where the highly accurate empirical PEC $U_X^{emp}$ of the ground $X^1\Sigma^+$ state was borrowed from Ref.~\onlinecite{Pashov2007}.

All electronic structure calculations were performed in a wide range of internuclear distances on the density grid by means of the MOLPRO program package~\cite{MOLPRO_brief}. The radial coupling matrix element $B_{12}(R)$ between the $1^1\Pi$ and $2^1\Pi$ states was evaluated by three points finite-difference method.

The details of the computational procedure used can be found elsewhere~\cite{Ab_KRb_2016}. Briefly, the inner core shell of both potassium and rubidium atoms was replaced by energy-consistent non-empirical effective core potentials~\cite{Lim05} (ECP), leaving 9 outer shell (8 sub-valence plus 1 valence) electrons for explicit treatment. The relevant spin-averaged and spin-orbit Gaussian basis sets used for each atom (ECP10MDF for K and ECP28MDF for Rb, respectively) were taken from the above reference. The optimized molecular orbitals were constructed from the solutions of the state-averaged complete active space self-consistent field problem for all 18 electrons on the lowest (1-10)$^{1,3}\Sigma^+$ and (1-5)$^{1,3}\Pi$ electronic states taken with equal weights~\cite{Werner85}. The dynamical correlation was introduced via the internally contracted multi-reference configuration interaction (MRCI) method~\cite{Knowles92} which was applied for only two valence electrons keeping the remaining 16 sub-valence electrons frozen. The $l$-independent core-polarization potentials (CPPs) of both atoms (see Table~\ref{CPP}) were employed to implicitly account for the residual core-valence correlation effects~\cite{Lim06}. The corresponding CPP cut-off radii of both atoms were adjusted to reproduce the experimental fine-structure splitting of the lowest excited K(4$^2$P) and Rb(5$^2$P) states~\cite{Atom}.

\begin{table}[h]
\small
\caption{The static dipole polarizability~\cite{AtomPol} of the cation and its cut-off radii implemented in the CPP potentials of the K and Rb atoms. All parameters in $a.u.$.}\label{CPP}
\begin{tabular}{ccc}
\hline\hline
  & $\alpha_{core}$ & $r_{cut-off}$\\
\hline
K  & 5.354 & 0.247\\
Rb & 9.096 & 0.379\\
\hline
\end{tabular}
\end{table}

The resulting MRCI wave functions were used to evaluate the permanent dipole functions $d_{1,2}(R)$ of  adiabatic $1^1\Pi$ and $2^1\Pi$ states as well as the corresponding $1^1\Pi-2^1\Pi$  transition dipole moment $d_{12}(R)$.
The adiabatic matrix elements were transformed to the relevant diabatic moments $\mu_{1,2}(R)$, $\mu_{12}(R)$ as
\begin{eqnarray}\label{ad_di_mu}
\mu_1 &=& \cos^2\theta d_1 + \sin^2\theta d_2 -\sin 2\theta d_{12}\nonumber\\
\mu_2 &=& \sin^2\theta d_1 + \cos^2\theta d_2 +\sin 2\theta d_{12}\nonumber\\
\mu_{12} &=& \cos 2\theta d_{12}+\sin 2\theta |d_1-d_2|/2
\end{eqnarray}

Finally, we have calculated spin-orbit $\xi_{ij}(R)$ ($j\in (1,2)^3\Pi;(2,3)^3\Sigma^+$) and angular $L^{\pm}_{ij}(R)$ ($j\in (1-3)^1\Sigma^+$) coupling matrix elements as well as transition dipole moments $d_{ij}(R)$ ($j\in (1,2)^1\Sigma^+$) for adiabatic $i\in 1^1\Pi,2^1\Pi$ states. The resulting adiabatic matrix elements $W_{ij}\in \xi_{ij},L^{\pm}_{ij}, d_{ij}$ were unitary transformed to their diabatic counterparts $\mathrm{W_{ij}}$ as
\begin{eqnarray}\label{ad_di_me}
\mathrm{W_{1j}} &=& \cos \theta W_{1j} + \sin \theta W_{2j}\nonumber\\
\mathrm{W_{2j}} &=&-\sin \theta W_{1j} + \cos \theta W_{2j}
\end{eqnarray}

\section{The coupled-channel deperturbation analysis of the $(1\sim 2)^1\Pi$ complex}\label{method}

In the framework of the rigorous coupled-channel (CC) deperturbation model~\cite{Bergeman, AB_LiAr_2005, Ab_KRb_2016}, the non-adiabatic rovibronic energy $E^{CC}$ of the $(1,2)^1\Pi$ complex is determined by the solution of the two coupled radial equations
\begin{eqnarray}\label{CC}
\left(-{\bf I}\frac{\hbar^2 d^2}{2\mu dR^2} + {\bf V}(R) -
{\bf I}E^{CC}\right)\mathbf{\Phi}(R) = 0\\
\mathbf{\Phi}(0)=\mathbf{\Phi}(\infty)=0,\nonumber
\end{eqnarray}
where $\mu$ is the reduced molecular mass, ${\bf I}$ is the identity matrix and ${\bf V}(R)$ is the symmetric matrix of the potential energy given by
\begin{eqnarray}\label{V12}
V_{1^1\Pi} = V_1;\quad V_{2^1\Pi} = V_2; \quad V_{1^1\Pi-2^1\Pi} = V_{12}
\end{eqnarray}
where the diabatic potentials $V_{1,2}(R)$ and electronic coupling matrix element $V_{12}(R)$ are the mass-invariant functions of internuclear distance $R$. The two-component vibrational eigenfunction $\mathbf{\Phi}$ in Eq.(\ref{CC}) is normalized for the bound states as $\sum_i P_i=1$, where $P_i=\langle\phi_i|\phi_i\rangle$ ($i\in 1^1\Pi, 2^1\Pi$) is the fractional partition of the level with the energy $E^{CC}$.

The diabatic matrix elements are related to the adiabatic PECs as
\begin{eqnarray}\label{adPECs}
U_{1,2} = (V_1 + V_2)/2\pm \sqrt{(V_1 - V_2)^2/4 + V_{12}^2}
\end{eqnarray}
The corresponding "effective" rotational constant $B^{CC}$ is defined as the expectation value
\begin{eqnarray}\label{Beff}
B^{CC}=\frac{\hbar^2}{2\mu}\sum_{i=1,2} \langle\phi_i|1/R^2|\phi_i\rangle
\end{eqnarray}

The rovibronic energies $E^{CC}$ and vibrational wave functions $\phi_i(R)$ were obtained through solving the CC equations (\ref{CC}) by the finite-difference boundary value method~\cite{duo}. The adaptive analytical mapping procedure~\cite{Meshkov08} was exploited to decrease the required number of grid points.

To perform a direct fit of the experimental data we represented the diabatic interatomic potentials and the relevant electronic coupling matrix element in their fully analytical forms.
In particular, the Morse/Long-Range(MLR)~\cite{leroy2006, leroy2007, salami2007} function
\begin{eqnarray}\label{MLR}
V_2(R) \equiv U_{MLR} = [T^{dis}_2 - {\mathfrak D}_e] + \\
{\mathfrak D}_e \left [ 1 - \frac{u_{LR}(R)}{u_{LR}(R_e)} e^{-\beta(R)y^{eq}_p(R)}\right ]^2
\nonumber
\end{eqnarray}
is used to approximate the diabatic PEC of the $2^1\Pi$ state converging to the K(4$^2$S)+Rb(5$^2$P) dissociation threshold. The fixed parameter of the dissociation energy $T^{dis}_2$ involved in Eq.(\ref{MLR}) was determined as $T^{dis}_2={\mathfrak D}_e^X+E_{Rb(5^2{\rm P})}- E_{Rb(5^2{\rm S})}=16955.169$~cm$^{-1}$, where the experimental dissociation energy of the ground state~\cite{Wang} (taken at the hfs center-of-gravity) ${\mathfrak D}_e^X$ = 4217.822 cm$^{-1}$ and the corresponding non-relativistic energy of the $D$-lines of the Rb atom~\cite{Atom}.

The coefficient $\beta(R)$ in the MLR function
\begin{equation} \label{V_MLR_beta}
\beta_{MLR}(R)=y^{ref}_p\beta_{\infty}+\left[1-y^{ref}_p\right]\sum^{N}_{i=0}\beta_i\left[y_q^{ref}\right]^i,
\end{equation}
is the polynomial function of the reduced coordinates $y_{p,q}^{ref}$:
\begin{equation} \label{V_MLR_y}
 y_{p,q}^{ref} (R) = \frac{ R^{p,q} - R^{p,q}_{ref} }{ R^{p,q} + R^{p,q}_{ref} },
\end{equation}
where $R_{ref}$ is the reference distance and the parameters $q$ and $p$ are integers. The reduced variable $y^{eq}_{p}$ in Eq.(\ref{MLR}) is defined by Eq.(\ref{V_MLR_y}) where the $R_{ref}$ is substituted for the equilibrium distance $R_e$. The parameter $\beta_{\infty}$ is constrained to be $\ln\left\{2{\mathfrak D}_e/u_{LR}(R_e)\right\}$, where ${\mathfrak D}_e$ is the the well depth and $u_{LR}$ is the long-range potential
\begin{equation} \label{V_MLR_u_LR}
  u_{LR}(R) = \sum_{n=6,8} D_n\frac{C_n}{R^n}
\end{equation}
with fixed sets of the dispersion coefficients $C_n$ and the damping functions~\cite{leroy2009,leroy2011} $D_n$ :
\begin{equation} \label{Damp_func_Douketis}
D_n(R)=\left [1-\exp\left(-\frac{3.3(\rho \cdot R)}{n}-\frac{0.423(\rho \cdot R)^2}{\sqrt{n}}\right)\right ]^{n-1},
\end{equation}
where the scaling parameter $\rho=0.461$~\cite{leroy2011}.

To approximate the diabatic PEC of the $1^1\Pi$ state converging to the K(4$^2$P)+Rb(5$^2$S) dissociation threshold we used the double-exponential/long-range (DELR) potential~\cite{huang2003}:
\begin{eqnarray}\label{DELR}
V_1(R)\equiv U_{DELR} = T^{dis}_1 - u_{LR}(R) + \\
A e^{-2\beta(R)(R-R_e)} - B e^{-\beta(R)(R-R_e)}
\nonumber
\end{eqnarray}
which allowed us to represent a rotationless barrier above the third asymptote at distances $R > R_e$ (see, Fig.~\ref{PECs_KRb}). The dissociation energy $T^{dis}_1={\mathfrak D}_e^X+E_{K(4^2{\rm P})}- E_{K(4^2{\rm S})}=17241.481$~cm$^{-1}$ was fixed during the fit. The non-relativistic energy of the $D$-lines of the K atom was taken from Ref.~\onlinecite{Atom}.

The exponent coefficient $\beta(R)$ of the DELR potential was defined as
\begin{equation}\label{betaED}
\beta_{DELR}(R) =\sum_{i=0}^{N} \beta_{i}\left[y_q^{ref}\right]^i
\end{equation}
while the pre-exponential coefficients $A$ and $B$ were determined from the conditions $U_{DELR}(R_e) = 0$ and $d U_{DELR}/dR|_{R=R_e}=0$ which lead to
\begin{eqnarray}\label{ABDELR}
A &=& {\mathfrak D}_e - u_{LR}(R_e) - u^{\prime}_{LR}(R_e)/\beta_{DELR}(R_e)\\
B &=& {\mathfrak D}_e - u_{LR}(R_e) + A\nonumber
\end{eqnarray}
where $u^{\prime}_{LR}\equiv d u_{LR}/dR$.

Finally, the electronic coupling matrix element $V_{12}^{emp}(R)$ between the diabatic $1^1\Pi$ and $2^1\Pi$ states was represented by the polynomial:
\begin{eqnarray}\label{V12emp}
V_{12}^{emp}(R)=(1-y_q^{ref}) \sum_{i=0}^N \beta_{i}\left[y_q^{ref}\right]^i
\end{eqnarray}

The optimal parameters of the MLR and DELR potentials as well as electronic coupling matrix element were determined simultaneously in the framework of the weighted nonlinear least-squared fitting (NLSF) procedure:
\begin{eqnarray}\label{chisexpab}
\chi^2&=&\sum^{N^{exp}}_{j=1}\left(E^{exp}_j-E^{CC}_j)/\sigma_j^{exp}\right)^2\\
&+&\sum_{j=1}^{N^{ab}}\left([V^{ab}_1(R_j)-U_{DELR}(R_j)]/\sigma_j^{ab}\right)^2\nonumber\\
&+&\sum_{j=1}^{N^{ab}}\left([V^{ab}_2(R_j)-U_{MLR}(R_j)]/\sigma_j^{ab}\right)^2\nonumber\\
&+&\sum_{j=1}^{N^{ab}}\left([V^{ab}_{12}(R_j)-V^{emp}_{12}(R_j)]/\sigma_j^{ab}\right)^2\nonumber
\end{eqnarray}
where $E^{exp}_j$ denote the experimental term values of the $(1\sim 2)^1\Pi$ complex and the $\sigma^{exp}_j$-values mean their uncertainties. The $V^{ab}_{1,2}$ and $V^{ab}_{12}$ are the diabatic functions evaluated by Eq.(\ref{ad_di}) at the point $R_j$, and $\sigma^{ab}_j$ are their uncertainties obtained by averaging the present and preceding~\cite{rousseau2000} \emph{ab initio} curves. The theoretical curves were incorporated in the NLSF procedure in order to propagate the empirical functions outside of the experimental data region.

\section{Results and discussion}\label{discussion}

\subsection{\emph{Ab initio} data}

\begin{figure}[h!]
\centering
\includegraphics[scale=0.4]{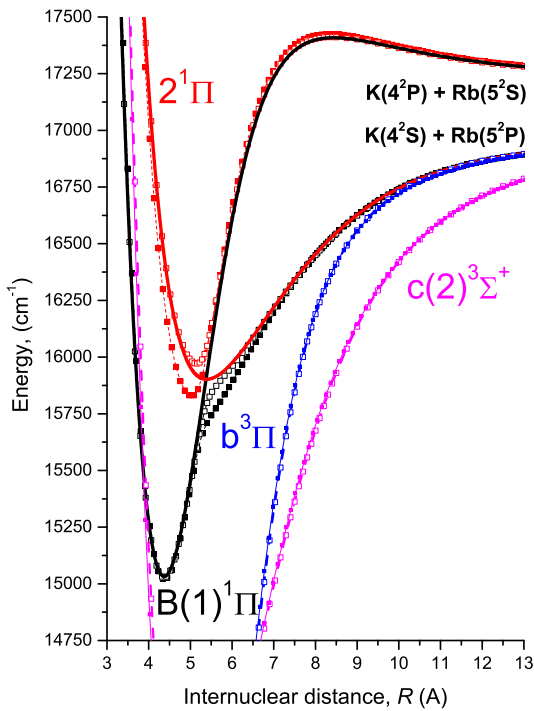}
  \caption{The "difference-based" adiabatic PECs obtained from the present (open symbols) and preceding~\cite{rousseau2000} (solid symbols) \emph{ab initio} calculations. The solid lines denote the corresponding diabatic \emph{ab initio} PECs of the $1^1\Pi$ and $2^1\Pi$ states.}
  \label{PECs_KRb_ab}
\end{figure}

The resulting "difference-based" PECs obtained by Eq.(\ref{difbased}) for adiabatic $B(1)^1\Pi$ and $2^1\Pi$ states from the present and preceding~\cite{rousseau2000} \emph{ab initio} calculations demonstrates overall good agreement as can be seen in Fig.~\ref{PECs_KRb_ab}. The most significant deviations are observed in the vicinity of the avoided crossing point $R_c\approx 5.36$~(\AA). The corresponding diabatic $V_1(R)$ and $V_2(R)$ PECs obtained by the unitary transformation (\ref{ad_di}) are depicted as well.

\begin{figure}[h!]
\centering
\includegraphics[scale=0.4]{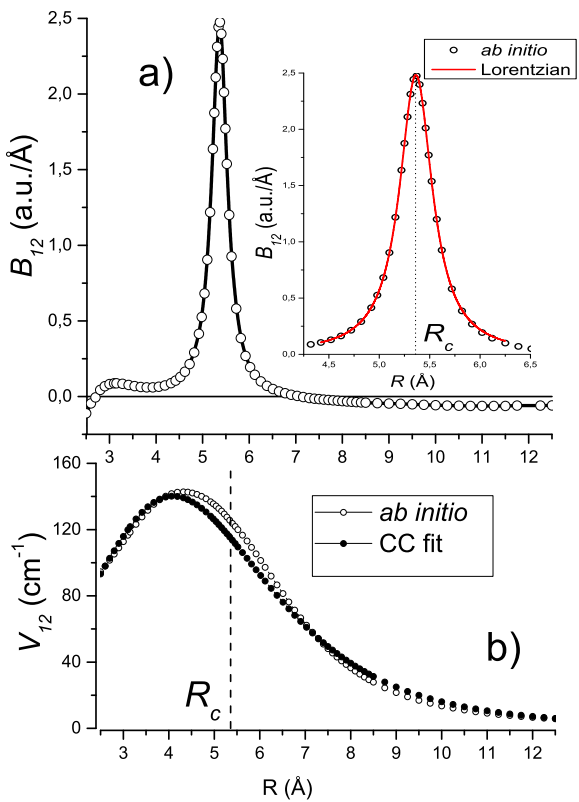}
\caption{(\textbf{a}) The radial coupling electronic matrix element $B_{12}(R)$ calculated between adiabatic $1^1\Pi$ and $2^1\Pi$ states. On the inset the red line depicts the fitting Lorentz curve (\ref{B12FIT}) with the parameters $R_c=5.36$ and $w=0.4036$ (both in ~\AA). (\textbf{b}) The \emph{ab initio} and empirical electronic coupling $V_{12}(R)$ function between diabatic $1^1\Pi$ and $2^1\Pi$ states.}
\label{B12_KRb_ab}
\end{figure}

The \emph{ab initio} radial $B_{12}(R)$ and electronic $V_{12}(R)$ coupling matrix elements are given on Fig.~\ref{B12_KRb_ab}a and b, respectively. The inset demonstrates that the simplest two-parameters Lorentz curve (\ref{B12FIT}) perfectly fits a peak of the \emph{ab initio} $B_{12}(R)$ function.

\begin{figure}[h!]
\centering
\includegraphics[scale=0.4]{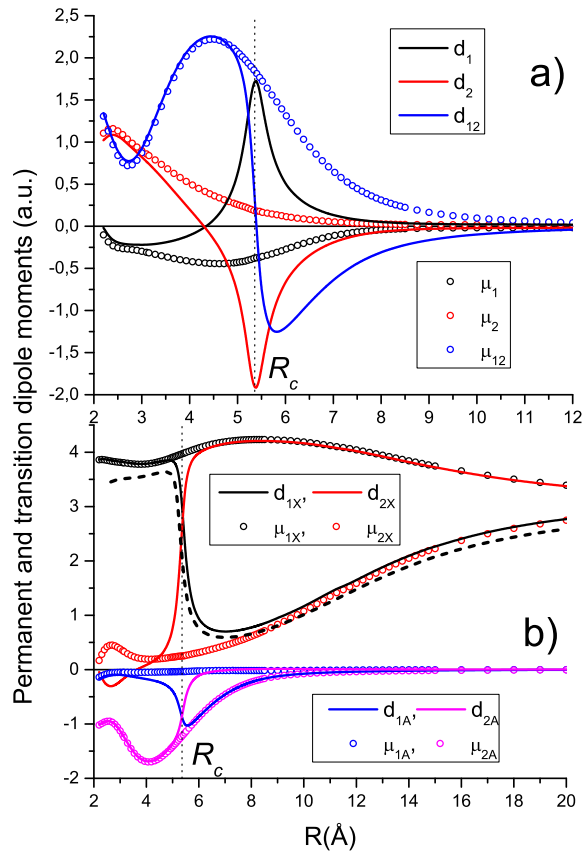}
  \caption{(\textbf{a}) The adiabatic and diabatic \emph{ab initio} permanent dipole moments of the $1^1\Pi$ and $2^1\Pi$ states as well as the corresponding $1^1\Pi-2^1\Pi$ transition moment. (\textbf{b}) The \emph{ab initio} TDM  functions between the $(1,2)^1\Pi$ and $X,A^1\Sigma^+$ states. The dashed line denotes the adiabatic $d_{B^1\Pi-X^1\Sigma^+}(R)$ function from Ref.~\onlinecite{Beuc2006}.}
  \label{d12_KRb_ab}
\end{figure}

The permanent dipole moments of the $1^1\Pi$ and $2^1\Pi$ states as well as the corresponding $1^1\Pi-2^1\Pi$ transition dipole moment are given on Fig.~\ref{d12_KRb_ab}a. The adiabatic functions $d_{1,2}(R)$, $d_{12}(R)$ were obtained during the electronic structure calculations while the diabatic moments $\mu_{1,2}(R)$, $\mu_{12}(R)$ were evaluated according to Eq.(\ref{ad_di_mu}). As follows from the charge density of the electronic wavefunctions of the twin $(1,2)^1\Pi$ states~\cite{leininger1995}, the $d_{1,2}(R)$ functions have a "mirror" $R$-dependance $d_1\approx -d_2$ with a sharp global extremum about $\pm 2$ $a.u.$ located near the point $R_c$. It should be noticed,  that the adiabatic transition moment $d_{12}$ becomes zero at the same point. In contrast to sharp adiabatic functions, their diabatic counterparts demonstrate rather smooth $R$-behavior. Furthermore, the absolute magnitudes of the diabatic $|\mu_{1,2}(R)|$ functions significantly decrease at intermediate internuclear distances.

The adiabatic $(1,2)^1\Pi-(X,A)^1\Sigma^+$ transition dipole moments are depicted on Fig.~\ref{d12_KRb_ab}b along with the diabatic moments evaluated by Eq.(\ref{ad_di_me}). A good agreement of the present $B(1)^1\Pi- X^1\Sigma^+$ transition moment and the preceding estimate~\cite{Beuc2006} is observed. The Fig.~\ref{d12_KRb_ab}b also shows that the diabatic $1^1\Pi-A^1\Sigma^+$ and $2^1\Pi\to X^1\Sigma^+$ moments are very small at short and intermediate $R$-distances.

The diabatic $1^1\Pi\to X^1\Sigma^+$ transition moment $\mu_{1X}(R)$ was used to evaluate a radiative lifetime for the lowest vibrational levels of the $1^1\Pi$ state by the approximate sum rule~\cite{PupyshevCPL94}:
\begin{eqnarray}\label{tausum}
\frac{1}{\tau_{1^1\Pi}}\approx\frac{8{\pi }^2}{3\hbar {\epsilon_0}}
\langle\phi_1^{J^{\prime}}|[\Delta V_{1X}]^3[\mu_{1X}]^2|\phi_1^{J^{\prime}}\rangle
\end{eqnarray}
where $\Delta V_{1X}(R)=V_1(R)-U_X(R)$ is the difference of the diabatic PEC of the $1^1\Pi$ state and ground state PEC. The resulting $\tau=11.3$ $ns$ predicted for the $B^1\Pi(v_1^{\prime}=2,J^{\prime}=41)$ level is remarkably close to its experimental counterpart~\cite{okada1996} of 11.6 $ns$. It should be noted, that the contribution of the $1^1\Pi - A^1\Sigma^+$ transition into the $\tau_{1^1\Pi}$-estimate could be neglected since $|\mu_{1A}|\ll |\mu_{1X}|$ and $|\Delta V_{1A}|\ll |\Delta V_{1X}|$.

\begin{figure}[h!]
\centering
\includegraphics[scale=0.4]{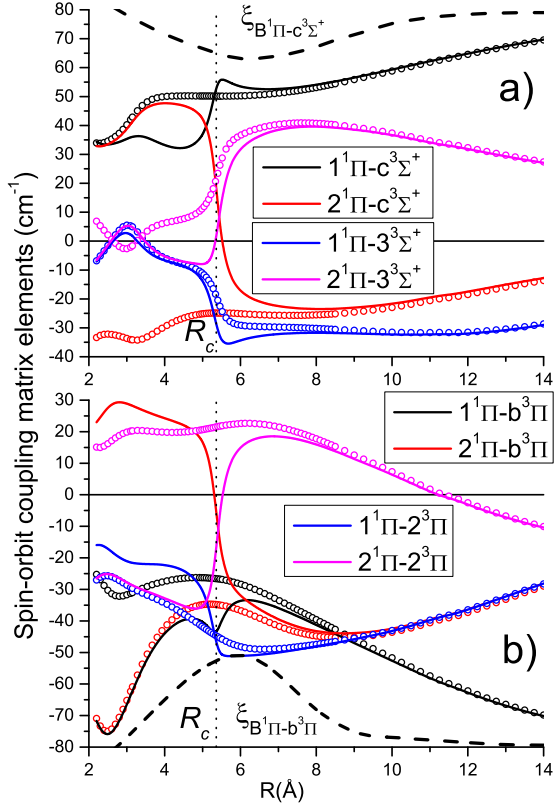}
  \caption{The adiabatic (solid lines) and diabatic (open circles) \emph{ab initio} spin-orbit coupling matrix elements $\xi_{ij}(R)$ between the $(1,2)^1\Pi$ and $(2,3)^3\Sigma^+$ (\textbf{a}), $(1,2)^3\Pi$ (\textbf{b}) states. The dashed lines denote the adiabatic $\xi_{B^1\Pi-c^3\Sigma^+}(R)$ and $\xi_{B^1\Pi-b^3\Pi}(R)$ SOC functions borrowed from Ref.~\onlinecite{Kotochigova2009}.}
  \label{SO_KRb_ab}
\end{figure}

The resulting SOC matrix elements obtained during the present \emph{ab initio} calculations are depicted on Fig.~\ref{SO_KRb_ab}. As expected, the diabatization procedure provides a smooth $R$-behavior of most SOC functions. However, the diabatic $(1,2)^1\Pi-3^3\Sigma^+$ functions are still not smooth enough since the radial coupling of the $3^3\Sigma^+$ state with the higher $(n\ge 4)^3\Sigma^+$ states takes place. It should be also noted that the present $B^1\Pi-c^3\Sigma^+$ and $B^1\Pi-b^3\Pi$ functions deviate significantly at short and intermediate $R$-ranges from the preceding result~\cite{Kotochigova2009}, which has been used in modeling the optimal $a^3\Sigma^+\to B^1\Pi \sim c^3\Sigma^+\to X^1\Sigma^+$ STIRAP cycle~\cite{borsalino2014}.

\begin{figure}[h!]
\centering
\includegraphics[scale=0.4]{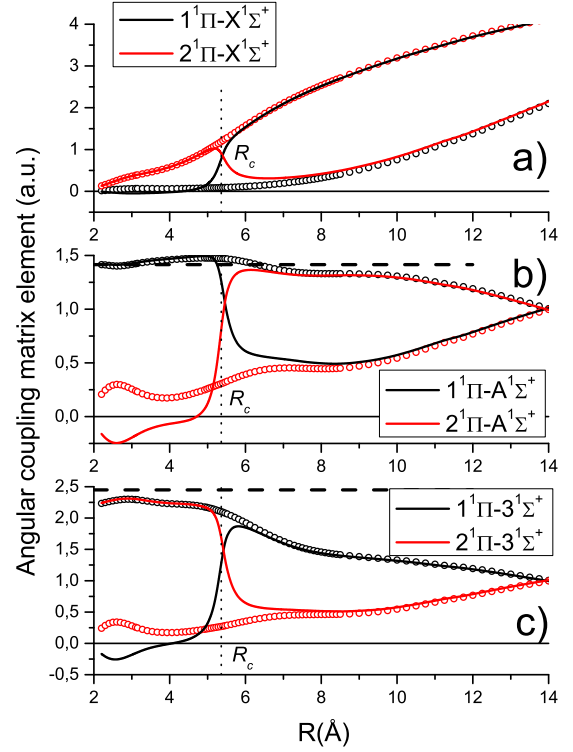}
  \caption{The adiabatic (solid lines) and diabatic (open circles) \emph{ab initio} angular coupling matrix elements $L^{\pm}_{ij}(R)$ calculated between the $(1,2)^1\Pi$ and $X^1\Sigma^+$ (\textbf{a}), $A^1\Sigma^+$ (\textbf{b}), $3^1\Sigma^+$ (\textbf{c}) states. The horizontal dashed lines correspond to $\sqrt{2}$ (\textbf{b}) and $\sqrt{6}$ (\textbf{c}) values, respectively.}
  \label{Lc_ab}
\end{figure}

The angular coupling matrix elements $L^{\pm}_{ij}(R)$ obtained for the $(1,2)^1\Pi-(1-3)^1\Sigma^+$ non-adiabatic transitions are presented on Fig.~\ref{Lc_ab}. It is seen that Van Vleck's \emph{pure precession} hypothesis~\cite{field} $L^{\pm}_{ij}\approx \sqrt{l(l+1)} = const$ works perfectly for the $1^1\Pi-A^1\Sigma^+$ pair with $l=1$. With $l=2$, it works fairly well for the $2^1\Pi-3^1\Sigma^+$ pair at short and intermediate distances.

Under \emph{unique perturber} approximation~\cite{field} the $q$-factors of the doubly degenerate $^1\Pi$ states are estimated as
\begin{eqnarray}\label{qfactor}
q_{^1\Pi}\approx \left (\frac{\hbar^2}{2\mu R_e^2}\right )^2\sum_{^1\Sigma^+}\frac{2l(l+1)}{T_e^{^1\Pi} -T_e^{^1\Sigma^+}}
\end{eqnarray}
yielding, for the $^{39}$K$^{85}$Rb diabatic states, $q_{1^1\Pi}\approx 3.2\times 10^{-5}$ and $q_{2^1\Pi}\approx 1.5\times 10^{-4}$ cm$^{-1}$, respectively. Unfortunately, there are no experimental $q$-values for comparison so far.

The resulting \emph{ab initio} potential energy curves, permanent and transition dipole moments as well as electronic, spin-orbit and angular coupling matrix elements are given in pointwise form in the Supplementary material~\cite{EPAPS}.

\subsection{CC deperturbation data}\label{empirical}

Experimental input data of the $1^1\Pi\sim 2^1\Pi$ complex used in the NLSF fitting procedure (\ref{chisexpab}) consists of (\emph{i}) the 110 original rovibronic termvalues~\cite{amiot2016} $E^{exp}_{v^{\prime}J^{\prime}}$ obtained during the $(3)^1\Pi\to (1,2)^1\Pi$ FTS LIF experiment~\cite{amiot2000} for the rotational levels $J^{\prime}\in [24,145]$ in the short range of vibrational quantum numbers $v_1^{\prime},v_2^{\prime}\in [0,6]$; (\emph{ii}) the reduced termvalues $E^{exp}_{v^{\prime}}$ obtained by the OODRPS measurements~\cite{okada1996, kasahara1999} for the $v_1^{\prime}\in [0,69]$ and $v_2^{\prime}\in [6,19]$ levels, respectively; (\emph{iii}) the rotationless ($J^{\prime}=0$) termvalues extracted from the MB experiment~\cite{kim2011} for $v^{\prime}_1\in [0,20]$ levels.

\begin{figure}[h!]
\centering
\includegraphics[scale=0.4]{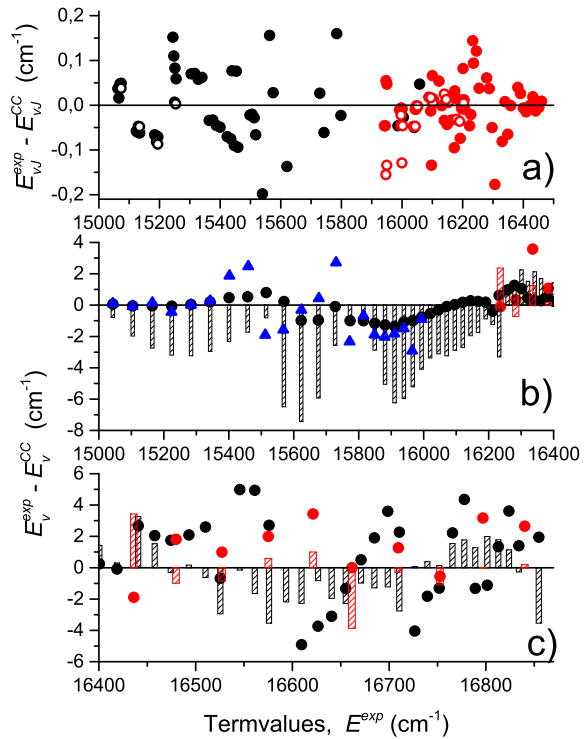}
  \caption{The residual of the experimental termvalues of the $1^1\Pi\sim 2^1\Pi$ complex and CC estimates (\ref{CC}). Black and red circles mark the $1^1\Pi$ and $2^1\Pi$ states, respectively. (\textbf{a}) Solid symbols correspond to the rovibronic termvalues~\cite{amiot2016} of $^{39}$K$^{85}$Rb isotopologue while open circles - $^{39}$K$^{87}$Rb. (\textbf{b-c}) The circles denote the OODRPS termvalues~\cite{okada1996, kasahara1999} while the triangles denote the MB data~\cite{kim2011}. The bars denotes the differences between the empirical terms and their estimates evaluated by the corresponding RKR potentials.}
  \label{term}
\end{figure}

All energies above correspond to the most abundant $^{39}$K$^{85}$Rb isotopologue while the 29 rovibronic termvalues~\cite{amiot2016} of $^{39}$K$^{87}$Rb held in reserve for confirmation of the mass-invariant properties of the fitting functions. The uncertainty $\sigma^{exp}$ of the raw rovibronic termvalues~\cite{amiot2016} was taken as 0.05 cm$^{-1}$ while the $\sigma^{exp}=2$ cm$^{-1}$ was adopted for the vibronic terms $E^{exp}_{v^{\prime}}$ since the difference of the empirical data~\cite{okada1996, kasahara1999} and Ref.~\onlinecite{kim2011} reach few reciprocal centimeters (see, Fig.~\ref{term}b).

The adjusted mass-invariant parameters of the DELR (\ref{DELR}) and MLR (\ref{MLR}) potentials obtained for the diabatic $1^1\Pi$ and $2^1\Pi$ states are presented on Table~\ref{PECfitparamDELR} and \ref{PECfitparamMLR}, respectively. The fitting parameters of the empiric $1^1\Pi\sim 2^1\Pi$ coupling function (\ref{V12emp}) are given on the Table~\ref{V12fitparam}. The resulting parameters are duplicated in non-truncated ASCII form in the Supplementary material~\cite{EPAPS}, where both experimental and CC term values along with their residuals (see Fig.\ref{term}) and fractional partitions are collected as well. The adiabatic PECs obtained by the transformation (\ref{adPECs}) from the empirical diabatic functions agree very well with the corresponding RKR potentials in the low energy region (see Fig.~\ref{PECs_KRb_emp}).

Fig.~\ref{term}a demonstrates that the present deperturbation model allows one to reproduce the most experimental rovibronic termvalues~\cite{amiot2016} of the $^{39}$K$^{85}$Rb isotopologue and to predict the $E^{exp}_{v^{\prime}J^{\prime}}$-values of the $^{39}$K$^{87}$Rb isotopologue with an uncertainty close to 0.05 cm$^{-1}$. However, there are several pronounced deviations of the experimental termvalues corresponding to the particular rotational levels of $v^{\prime}_1=4,5$ vibrational states from the CC estimates (see the Supplementary material~\cite{EPAPS} for details). The observed outliers are attributed to the local SOC effect with the lower lying $c(2)^3\Sigma^+$ state (see, Fig.~\ref{PECs_KRb_ab}).

\begin{figure}[h!]
\centering
\includegraphics[scale=0.4]{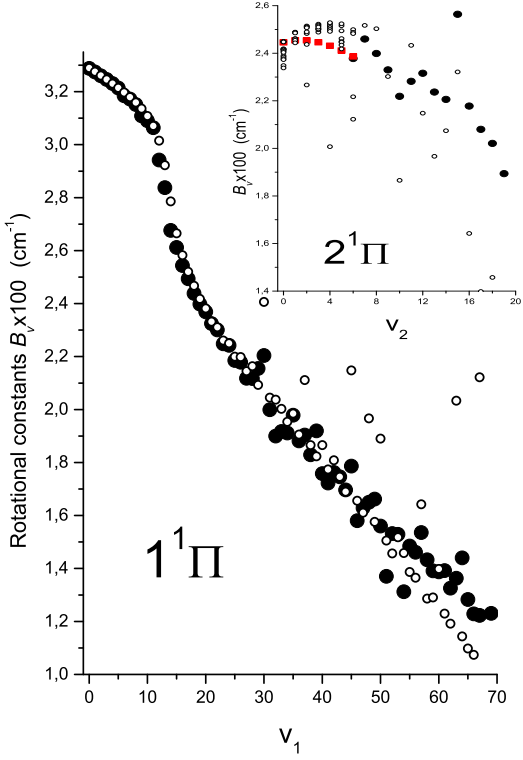}
  \caption{Comparison of the empirical (solid circles) rotational constants $B_{v^{\prime}}^{exp}$ of the $1^1\Pi\sim 2^1\Pi$ complex~\cite{okada1996, kasahara1999} and theoretical $B_{v^{\prime}}^{CC}$ estimates (open circles) derived by Eq.(\ref{Beff}). Red squares on the inset denote $B_{v^{\prime}}^{exp}$-values calculated for the $v_2^{\prime}\le 6$ levels by the Dunham constants~\cite{amiot2000}.}
  \label{rotconst}
\end{figure}

The CC diabatic model also improves the representation of the vibronic $E^{exp}_{v^{\prime}}$ termvalues~\cite{okada1996, kasahara1999, kim2011} up to the excitation energies $\sim 16400$ cm$^{-1}$ (see Fig.~\ref{term}b). Overall good agreement of the calculated $B_{v^{\prime}}^{CC}$ (\ref{Beff}) and empirical $B_{v^{\prime}}^{exp}$ rotational constants is observed on Fig.~\ref{rotconst} for the vibrational $v_1^{\prime}\le 40$ and $v_2^{\prime}\le 6$ levels.

\begin{figure}[h!]
\centering
\includegraphics[scale=0.4]{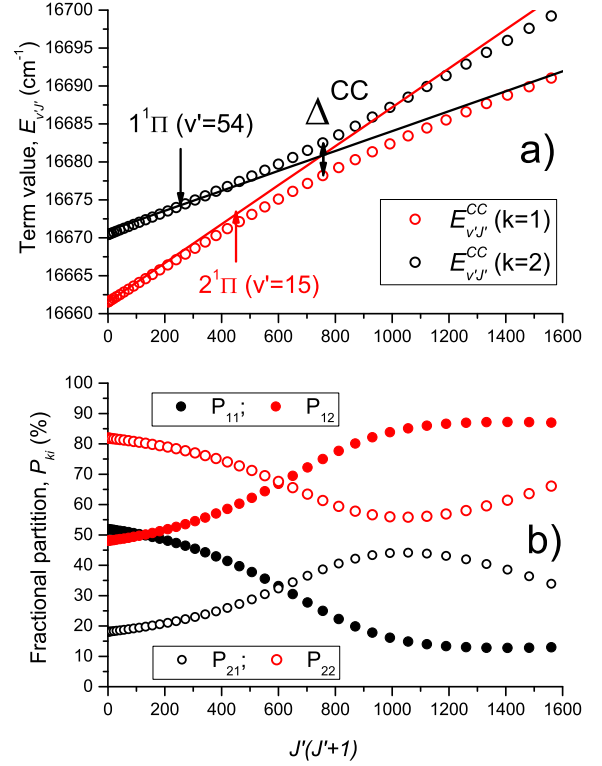}
  \caption{(\textbf{a}) The rovibronic termvalues $E^{CC}_{v^{\prime}J^{\prime}}$ predicted for the $1^1\Pi(v^{\prime}=54)\sim 2^1\Pi(v^{\prime}=15)$ levels under the present CC deperturbation model. The straight lines denotes the adiabatic energies calculated as $E^{exp}_{v^{\prime}}+B^{exp}_{v^{\prime}}\times J^{\prime}(J^{\prime}+1)$ by the "effective" band constants~\cite{kasahara1999}. The symbol $\Delta^{CC}$ means the minimal distance predicted at $J^{\prime}=27$. (\textbf{b}) The fraction partition $P_{ki}$ of the CC eigenfunctions corresponding to termvalues above.}
  \label{termcross}
\end{figure}

To test extrapolation possibilities of the deperturbation model we have calculated rovibronic termvalues $E^{CC}_{v^{\prime}J^{\prime}}$ for a pair of the closely lying $1^1\Pi(v^{\prime}_1=54)\sim 2^1\Pi(v^{\prime}_2=15)$ levels of the $(1\sim 2)^1\Pi$ complex (see Fig.~\ref{termcross}) experimentally studied in Ref.~\onlinecite{kasahara1999}. The vibrational numbering used above corresponds to the adiabatic representation of the $1^1\Pi\sim 2^1\Pi$ complex applied for the assignment of the OODRPS $X^1\Sigma^+\to 1^1\Pi\sim 2^1\Pi$ spectra~\cite{kasahara1999}. The calculated $E^{CC}_{v^{\prime}J^{\prime}}$ positions are found to be in a good agreement with their experimental counterparts. In particular, the minimal distance $\Delta^{CC}=4.3$ cm$^{-1}$ predicted at $J^{\prime}=27$ is remarkably close to the empirical estimate $\Delta^{exp}=2\times 2.2$ cm$^{-1}$ obtained in Ref.~\onlinecite{kasahara1999}. The fraction partition $P_{ki}\equiv\langle\phi_{ki}^{J^{\prime}}|\phi_{ki}^{J^{\prime}}\rangle$ of the CC vibrational eigenfunctions highlights a strong dependance of admixture of states on the rotational quantum number in the interval $J^{\prime}\in [11,35]$.

The divergence of the present CC estimates and empirical band constants generally increases as the vibrational excitation increases (see Fig.~\ref{term}c). The same effect takes places in the empirical adiabatic PECs and RKR potentials. It can be attributed to the monotonically growing SO coupling with the $b^3\Pi$ state correlated with the same dissociation limit (see, Fig.~\ref{PECs_KRb_ab}). Furthermore, the high vibrational $1^1\Pi(v_1^{\prime}\ge 63)$ levels lying just above the fine 4$^2$S$_{1/2}$(K)+5$^2$P$_{1/2}$(Rb) asymptotic undergo a predissociation effect~\cite{kasahara1999}.

Thus, raw experimental termvalues corresponding to high $v^{\prime},J^{\prime}$-levels of the $(1\sim 2)^1\Pi$ complex would be certainly useful for the comprehensive deperturbation analysis since the "effective" band constants indispensably "absorb" the spin0orbit perturbation effect in the high energy region.

\begin{table}[h!]
\caption{The resulting mass-invariant parameters of the $U_{DELR}(R)$ potential (\ref{DELR}) obtained for the diabatic $1^1\Pi$ state.}
\label{PECfitparamDELR}
\begin{center}
\begin{tabular}{lr}
\hline \hline
\multicolumn{2}{c}{fitted}\\
\hline
${\mathfrak D}_e$, cm$^{-1}$     &  2214.757      \\
$R_e$  , \AA                     &  4.3715        \\
$A$    , cm$^{-1}$               &  -10280.36     \\
$B$    , cm$^{-1}$               &   4577.51      \\
$\beta_0$, \AA$^{-1}$            &  0.47730       \\
$\beta_1$, \AA$^{-1}$            &  0.22714       \\
$\beta_2$, \AA$^{-1}$            &  0.11574       \\
$\beta_3$, \AA$^{-1}$            & -0.04563       \\
$\beta_4$, \AA$^{-1}$            & -0.20030       \\
$\beta_5$, \AA$^{-1}$            &  0.19131       \\
$\beta_6$, \AA$^{-1}$            &  0.73518       \\
$\beta_7$, \AA$^{-1}$            & -0.15280       \\
$\beta_8$, \AA$^{-1}$            & -0.72517       \\
\hline
\multicolumn{2}{c}{fixed}\\
\hline
$q$                              &  3             \\
$R_{ref}$,~\AA                   &  5.6781        \\
$T^{dis}$,~ cm$^{-1}$            &  17241.481      \\
$C_6$, cm$^{-1}\cdot$\AA$^6$     & $-2339\times 10^5$\\
$C_8$, cm$^{-1}\cdot$\AA$^8$     & $9536\times 10^5$ \\
\hline
\end{tabular}
\end{center}
\end{table}

\begin{table}[h!]
\caption{The resulting mass-invariant parameters of the $U_{MLR}(R)$ (\ref{MLR}) potential obtained for the diabatic $2^1\Pi$ state.}
\label{PECfitparamMLR}
\begin{center}
\begin{tabular}{lc}
\hline \hline
\multicolumn{2}{c}{fitted}\\
\hline
${\mathfrak D}_e$,~cm$^{-1}$ &   1111.961\\
$R_e$,~\AA                   &   5.2872  \\
$\beta_0$                    &  -0.92350 \\
$\beta_1$                    &   0.09744 \\
$\beta_2$                    &   0.09023 \\
$\beta_3$                    &  -0.71368 \\
$\beta_4$                    &  -1.08251 \\
$\beta_5$                    &   0.12502 \\
$\beta_6$                    &   0.35828 \\
\hline
\multicolumn{2}{c}{fixed}\\
\hline
$q$                        &   4                     \\
$p$                        &   4                     \\
$R_{ref}$,~\AA             &  6.8511                 \\
$T^{dis}$,~ cm$^{-1}$      &  16955.169              \\
$C_6$,~cm$^{-1}\cdot$\AA$^6$    &  $3025\times 10^5$\\
$C_8$,~cm$^{-1}\cdot$\AA$^8$    &  $7875\times 10^5$\\
\hline
\end{tabular}
\end{center}
\end{table}

\begin{table}[h!]
\caption{The resulting mass-invariant parameters of the electronic $1^1\Pi\sim 2^1\Pi$ coupling $V_{12}^{emp}(R)$ function (\ref{V12emp}).}
\label{V12fitparam}
\begin{center}
\begin{tabular}{lc}
\hline \hline
\multicolumn{2}{c}{fitted}\\
\hline
$\beta_0$                  &  114.87  \\
$\beta_1$                  & -2.4087  \\
$\beta_2$                  & -98.392  \\
\hline
\multicolumn{2}{c}{fixed}\\
\hline
$q$                        &   3      \\
$R_{ref}$,~\AA             &   5.36   \\
\hline
\end{tabular}
\end{center}
\end{table}

\subsection{Intensity anomalies of the $X^1\Sigma^+(v^{\prime\prime}=12)\to 1^1\Pi(v^{\prime}=54)\sim 2^1\Pi(v^{\prime}=15)$ transition}
\label{anomalies}

The present deperturbation model (\ref{CC}) combined with diabatic $(1,2)^1\Pi - X^1\Sigma^+$ transition moments $\mu_{1X}(R)$, $\mu_{2X}(R)$ (see Fig.~\ref{d12_KRb_ab}b) was used to elucidate the "abnormal" intensity distribution observed for the $X^1\Sigma^+(v_X^{\prime\prime}=12,J^{\prime\prime}=J^{\prime}+1)\to 1^1\Pi(v_1^{\prime}=54, J^{\prime})\sim 2^1\Pi(v_2^{\prime}=15,J^{\prime})$ rovibronic transition~\cite{kasahara1999}.

\begin{figure}[h!]
\centering
\includegraphics[scale=0.4]{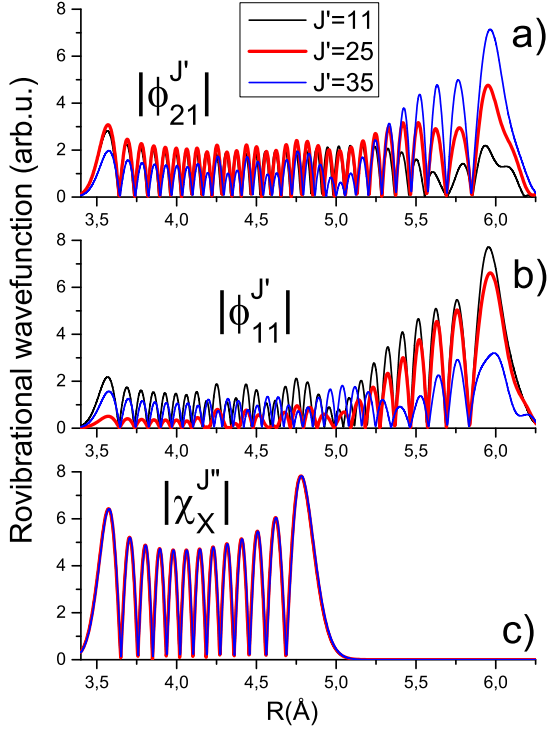}
  \caption{The nodal structure of vibrational wavefunctions calculated for the particular rotational $J^{\prime}=J{^{\prime\prime}}-1$ levels of the upper $v_1^{\prime}=54$ (\textbf{a}), $v_2^{\prime}=15$ (\textbf{b}) and ground $v_X^{\prime\prime}=12$ (\textbf{c}) vibrational states.}
  \label{wavefunction}
\end{figure}

The absorbtion intensities from the ground $X^1\Sigma^+$ state to a pair of adjoining levels of the $(1\sim 2)^1\Pi$ complex were evaluated according to the relation:
\begin{eqnarray}\label{Iteninterfer}
I^{CC}_{k^1\Pi-X^1\Sigma^+} &\sim& |\rm{M_{kX}}|^2\\
&=&|\langle\phi_{k1}^{J^{\prime}}|\mu_{1X}|\chi_X^{J^{\prime\prime}}\rangle|^2 + |\langle\phi_{k2}^{J^{\prime}}|\mu_{2X}|\chi_X^{J^{\prime\prime}}\rangle|^2\nonumber\\
&+&2\langle\phi_{k1}^{J^{\prime}}|\mu_{1X}|\chi_X^{J^{\prime\prime}}\rangle\langle\phi_{k2}^{J^{\prime}}|\mu_{2X}|\chi_X^{J^{\prime\prime}}\rangle\nonumber
\end{eqnarray}
where $k = 1, 2$ is the index of the states represented on Fig.~\ref{termcross}, $\phi_{k1}^{J^{\prime}}(R)$, $\phi_{k2}^{J^{\prime}}(R)$ are the CC rovibrational wavefunctions and $\chi_X^{J^{\prime\prime}}(R)$ are the vibrational wavefunctions of the ground $X$-state (see Fig.~\ref{wavefunction}) calculated with the highly accurate empirical potential~\cite{Pashov2007}.

\begin{figure}[h!]
\centering
\includegraphics[scale=0.4]{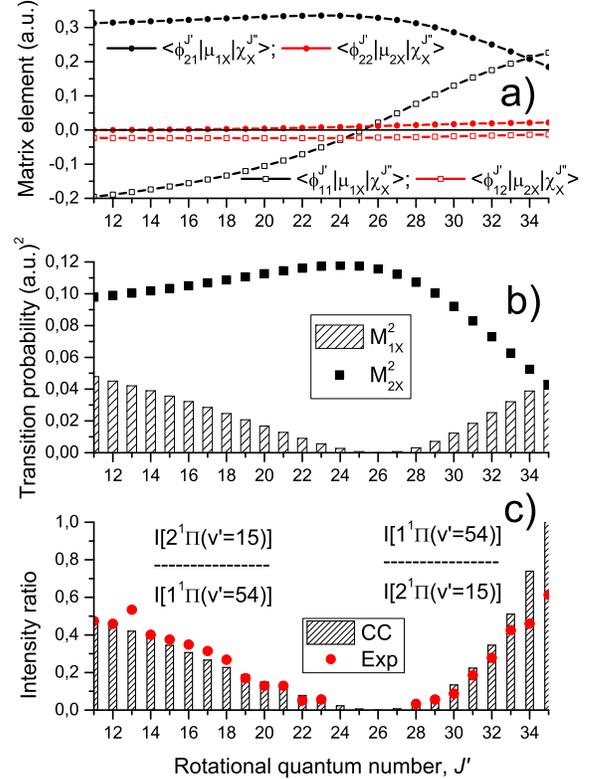}
  \caption{(\textbf{a}) The rovibronic transition matrix elements calculated for the $v_X^{\prime\prime}=12\to v_1^{\prime}=54\sim v_2^{\prime}=15$ transition; (\textbf{b}) The total $X^1\Sigma^+\to 1^1\Pi\sim 2^1\Pi$ transition probabilities; (\textbf{c}) Comparison of the theoretical (\ref{Iteninterfer}) and experimental~\cite{kasahara1999} intensity ratios. The experimental points were digitized from Fig.8 of Ref.~\onlinecite{kasahara1999}.}
  \label{intensity}
\end{figure}

The rovibronic $\langle\phi_{k1}^{J^{\prime}}|\mu_{1X}|\chi_X^{J^{\prime\prime}}\rangle$ and $\langle\phi_{k2}^{J^{\prime}}|\mu_{2X}|\chi_X^{J^{\prime\prime}}\rangle$ matrix elements calculated for the $X^1\Sigma^+(v_X^{\prime\prime}=12)\to 1^1\Pi(v_1^{\prime}=54)\sim 2^1\Pi(v_2^{\prime}=15)$ transitions are given on Fig.~\ref{intensity}a. It is seen that the $\langle\phi_{k2}^{J^{\prime}}|\mu_{2X}| \chi_X^{J^{\prime\prime}}\rangle$ terms give a negligible contribution to the total $|\rm{M_{kX}}|^2$ transition probability since $|\mu_{2X}|\ll |\mu_{1X}|$ (see Fig.~\ref{d12_KRb_ab}b). Furthermore, $\langle\phi_{21}^{J^{\prime}}|\mu_{1X}|\chi_X^{J^{\prime\prime}}\rangle$ matrix elements demonstrate abnormally strong $J^{\prime}$-dependance, and they accidentally become very small in the vicinity of $J^{\prime}\approx 25$ due to the interference effect taking place in the overlap integral of the upper and ground rovibrotional wavefunctions.

\section{Concluding remarks}

We performed the diabatization of the twin $(1,2)^1\Pi$ states of KRb based on an \textit{ab initio} electronic structure calculation and the direct coupled-channel treatment of experimental term values of the $(1\sim 2)^1\Pi$ complex. The present CC deperturbation model, based on a diabatic representation, provides the almost spectroscopic (experimental) accuracy of the approximation. The empirical PECs and electronic coupling function, along with \emph{ab initio} spin-orbit and angular coupling matrix elements, could be utilized in further deperturbation analysis carried out in the framework of both adiabatic and diabatic approximation. The diabatic transition dipole moments are appropriated for radiative property estimates.

\section*{Acknowledgments}
Authors are indebted to Claude Amiot for providing the raw rovibronic termvalues of the $(1\sim2)^1\Pi$ complex and Ekaterina Bormotova for fruitful discussion. The work was partly supported by RFBR grant No. 16-03-00529a.

\section*{References}

\bibliography{short,mybibfile}

\end{document}